\documentclass[twocolumn,showpacs,showkeys,preprintnumbers,amsmath,amssymb]{revtex4}

\usepackage{graphicx}

\usepackage{dcolumn}
\usepackage{bm}

\providecommand{\beqa}{\begin{eqnarray}}
 \providecommand{\bf}{\mathbf}
 \providecommand{\rm}{\mathrm}
\providecommand{\eeqa}{\end{eqnarray}}

\def\Z2{{\mathbf{Z}_2}}

\begin{document}


\title{Gravity Waves from Chain Inflation}

\author{Amjad Ashoorioon}
\email{amjad@umich.edu}

\author{Katherine Freese}
\email{ktfreese@umich.edu}

\affiliation{Michigan Center for Theoretical Physics, University of
Michigan, Ann Arbor, Michigan 48109-1040, USA}

\date{\today}

\begin{abstract}

Chain inflation proceeds through a series of first order phase
transitions, which can release considerable gravitational waves
(GW). We demonstrate that bubble collisions can leave an observable
signature for future high-frequency probes of GWs, such as advanced
LIGO, LISA and BBO. A "smoking gun" for chain inflation would be
wiggles in the spectrum (and consequently  in the tensor spectral
index) due to the multiple phase transitions. The spectrum could
also be distinguished from a single first order phase transition by
a small difference in the amplitude at low frequency.  A second
origin of GWs in chain inflation are tensor modes from quantum
fluctuations; these GW can dominate and be observed on large scales.
The consistency relation between scalar and tensor modes is
different for chain inflation than for standard rolling models and
is testable by Cosmic Microwave Background experiments. If inflation
happened through a series of rapid tunnelings in the string
landscape, future high frequency probes of GW can shed light on the
structure of the landscape.

\end{abstract}

\pacs{98.80.Cq}
\keywords{Chain Inflation, Gravity Wave, Bubble Collision}

\maketitle

Several experiments are underway to detect the stochastic
gravitational background from early universe cosmology. These
gravity waves (GWs) propagate almost freely throughout the entire
history of the universe and thus can be a direct source of
information about the universe at very early times. GW experiments
include the ground-based LIGO \cite{LIGO} and VIRGO \cite{VIRGO}
taking data now, and the upcoming space-based LISA \cite{LISA}. In
the more distant future are the proposed space-interferometers, BBO
\cite{BBO}, with possible launch within $20$ or $30$ years.

One  source of GWs is inflation  \cite{Guth:1980zm}, a quasi-exponential growth
phase of the universe which can explain its homogeneity and isotropy as well as
the generation of density perturbations required for structure formation.
There are two types of contributions to GW from inflation.  In slowly  rolling
models, quantum fluctuations of space-time lead to contributions proportional
to the height of the inflaton potential.  In tunneling models, there is an additional
 contribution due to bubble collisions of true vacuum bubbles at the end of
the phase transition \cite{Turner:1992tz,Kosowsky:1992vn,Kosowsky:1992rz}.
In this paper we consider the model of chain inflation, which gives rise to  both types of contributions.
Regarding inflation, on $\sim$ horizon scales, WMAP constrains the GW amplitude,
 $\Omega_{\mathrm{GW}} h^2\lesssim
10^{-15}$ \cite{Smith:2005mm}.  Future proposed cosmological
probes, e.g., CMBPOL, would be able to detect GW from inflation
at large scales, if the energy scale of inflation $\sim$ GUT scale
\cite{inf-work-group}.

In chain inflation \cite{Freese:2004vs}, the universe tunnels rapidly
through a series of first order phase transitions.  One can
imagine a multidimensional potential containing many minima of varying energy.
The universe starts out in a high-energy minimum, and then sequentially tunnels down
to bowls of ever lower energy until it reaches the bottom. During the time spent in any
one of these minima, the universe inflates by a fraction of an $e$-fold.  After many
hundreds of tunneling events, the universe has inflated by the 60 (or so)
$e$-folds required to resolve the cosmological problems.  In one variant of chain inflation,
the tunneling corresponds to quantized changes in four form fluxes \cite{Freese:2006fk} such as may be
found in the string landscape.  Other variants have been discussed in
\cite{Freese:2005kt,Chialva:2008zw}, but see
\cite{Ashoorioon:2008pj}. At each stage, the phase
transition is rapid enough that bubbles of true vacuum intersect one another and percolation is complete.  Thus
the failure of "old" inflation \cite{Guth:1980zm}  to reheat is avoided,
and "graceful exit" is achieved.

In the zero temperature limit, the nucleation rate $\Gamma$ per unit spacetime volume at a
phase transition
for producing bubbles of true vacuum in the sea of false vacuum through quantum tunneling is
$\Gamma(t) = A
e^{-S_E}$, where $S_E$ is the Euclidean action for the bounce
solution extrapolating between false and true vacua
\cite{Coleman:1977py} and $A$ is a determinantal factor. For a first order phase transition, with
Einstein gravity, it has been shown that the probability of a point
remaining in the false vacuum is given by
\begin{equation}\label{prob-desitter}
  p(t)\sim \exp(-\frac{4\pi}{3}\gamma H t),
\end{equation}
where $\gamma$ is defined as
\begin{equation}\label{gamma}
  \gamma\equiv\frac{\Gamma}{H^4}.
\end{equation}
Writing equation (\ref{prob-desitter}) as $\exp(-t/\tau)$, the
lifetime of the field in the false vacuum is
$\tau=\frac{3}{4\pi\gamma H}$. The number of e-foldings for the
tunneling event is
\begin{equation}\label{chi}
 \chi=\int Hdt\sim H\tau=\frac{3}{4\pi\gamma}.
\end{equation}
$\gamma$ has to be greater than critical value, $\gamma_c$, where
\cite{Turner:1992tz}
\begin{equation}\label{gamma_c}
 \gamma_c=9/4\pi,
\end{equation}
to achieve percolation and thermalization. This corresponds to
having an upper bound on the number of e-foldings that that can be
obtained in each stage of inflation:
\begin{equation}\label{chi-critical}
 \chi\leq \chi_c=\frac{1}{3}.
\end{equation}

The number of e-foldings  required for successful inflation
depends on the energy scale of inflation and
the reheating temperature. For a model with energy
scale  $\sim$ GUT scale, about $60$
e-foldings are required. This corresponds to having at least $\sim$
$200$ tunneling events for  GUT scale chain inflation.

A study of density perturbations from chain inflation has been done
by \cite{Chialva:2008zw,Feldstein:2006hm}; the former set of authors found that the
right perturbations to match data can be generated. In
this letter we calculate the gravity spectrum from chain inflation.
Since chain inflation proceeds through first order phase
transitions, one expects a considerable amount of GWs from chain
inflation.

The spectrum of gravity waves (GWs) from multi-bubble collisions at a {\it single} phase transition (PT)
 with energy difference $\epsilon$ between false and true vacua was
worked out numerically in \cite{Kosowsky:1992vn}.   In this paragraph we review their results,
and in the next section generalize to the multiple PT of chain inflation.  We note that
an example of  a single PT model which resolves
the graceful exit problem of old inflation by having time-dependent nucleation rates is in \cite{Adams:1991ma},\cite{linde}.
The authors of \cite{Kosowsky:1992vn}  found that the fraction of vacuum energy
released into gravity waves has the following spectrum:
\begin{equation}\label{GWprofile}
I(f) \equiv E_{GW}(f) / \epsilon=\left\{
\begin{array}{ll}
I_{\rm max} {\left(\frac{f}{f_{\rm max}}\right)}^{2.8}\quad f\leq  f_{\rm max} &\\
I_{\rm max} {\left(\frac{f}{f_{\rm max}}\right)}^{-1.8}\quad f\geq  f_{\rm max}. &
\end{array}
\right.
\end{equation}
The peak of the spectrum is at
\begin{equation}\label{fmax}
f_{\rm max}\simeq 0.2\beta,
\end{equation}
where
\begin{equation}\label{EGW}
I_{\rm max}\equiv \frac{E_{\rm GW}}{\epsilon} (f_{\max})\simeq 6\times 10^{-2}\left(\frac{H}{\beta}\right) ,
\end{equation}
\begin{equation}\label{beta}
\beta \equiv \frac{d\ln \Gamma(t)}{d t}=\frac{4\pi\gamma
H}{3}=\frac{H}{\chi}.
\end{equation}
Redshifting  the frequency and energy of gravitational radiation
as $a^{-1}$ and $a^{-4}$ respectively, at the current epoch the peak
frequency, $f_0$, and the corresponding logarithmic contribution of GWs to critical
density at the peak are given by \cite{Kosowsky:1992rz}:
\begin{eqnarray}\label{f0-omega0}
f_{0}\simeq  \frac{3\times 10^{-10}}{\chi}{\left(\frac{g_{\ast}}{100}\right)}^{1/6}\left({\frac{T_{\ast}}{1~{\rm GeV}}}\right),\\
\Omega_0(f_0) h^2\equiv \frac{1}{2\pi\rho_c} \frac{dE_{\rm GW}}{d\ln f} (f_0)
\simeq 10^{-6} {\chi^2} {\left(\frac{100}{g_{\ast}}\right)}^{1/3}.
\end{eqnarray}
Here $\Omega_0$ is the  GW energy density in units of the critical density $\rho_c$
required to close the universe and $h$ is today's Hubble constant in units of 100 km/s/Mpc.
$T_{\ast}$ is the temperature increase right after a single phase transition,
\begin{equation}\label{Tstar}
T_{\ast}={\left(\frac{30 \epsilon}{g_{\ast}\pi^2} \right)}^{1/4},
\end{equation}
and the total number of relativistic degrees of freedom at
temperature $T_{\ast}$ is taken to be $g_{\ast} \simeq 100$.
We note that in the past few years, the GW resulting from bubble collisions have been reexamined by
\cite{Huber:2007vva} and \cite{Caprini:2007xq}.  The former find
 that the fall-off at the low-frequency tail  $\sim f^{-1}$, rather than $f^{-1.8}$; this result leads to a smaller fall-off of GW amplitude and hence improves detectability.  To be conservative, we will mainly stick to the result of \cite{Kosowsky:1992vn} and compare with the consequences of \cite{Huber:2007vva} as needed.



{\bf Spectrum of GWs from Bubble Collisions:} We  now generalize these results to a series of PT to obtain GW from a series of bubble collisions in chain inflation. Let us assume that the total number of minima that the universe passes
through during chain inflation is $N$. As we will see its precise value will not
have any impact on our final result. Then we will have $N-1$ PTs
during the course of chain inflation. The energy difference between
two consecutive minima is assumed to be the constant value of
$\epsilon$. We also assume that $\gamma$ does not change and it
satisfies the condition $\gamma\geq\gamma_c$. Assuming instant
reheating after each PT, the temperature of the universe rises after each stage by
an amount $T_{\ast}$, where $T_{\ast}$ is given in equation (\ref{Tstar}).
The universe's temperature redshifts by a factor of $\exp(-\chi)$ during each
bout of chain inflation. Thus the temperature of the universe at the
end of chain inflation is
\begin{equation}\label{Tast1}
T_e=T_{\ast}(1+e^{-\chi}+\cdots+e^{-(N-1)\chi})=T_{\ast} \frac{\left(1-e^{-N\chi}\right)}{\left(1-e^{-\chi}\right)}
\end{equation}
For $N\chi\gg 1$, the final temperature of the universe is simplified to
\begin{equation}\label{Tast2}
T_e=\frac{T_{\ast}}{1-e^{-\chi}}.
\end{equation}
As stated above, every first order PT can release a substantial
amount of GWs from bubble collisions whose profile, peak frequency
and energy fraction at the peak frequency are given by eqs.
(\ref{GWprofile}), (\ref{fmax}) and (\ref{EGW}). After each PT
energy density in the gravitational radiation decreases like
$a^{-4}$, whereas the frequency redshifts like $a^{-1}$ \footnote{Notation: henceforth we will
suppress the subscript $0$ referring to values at the present epoch; e.g., $f_m$ is the
frequency today.}  The
peak frequency and fraction of critical density from GWs at the peak
frequency generated during the $m$-th PT redshifts as:
\begin{eqnarray}\label{fm1}
f_m&=&\left(\frac{a_{m}}{a_{e}}\right)\left(\frac{a_{e}}{a_0}\right) f_{\rm max},\\
\Omega h^2 (f_m)&
=&{\left(\frac{a_{m}}{a_{e}}\right)}^{4}{\left(\frac{a_{e}}{a_0}\right)}^{4}
\left.\frac{\epsilon}{2\pi\rho_c}\frac{d I(f)}{d\ln
f}\right|_{f=f_m},
\end{eqnarray}
where $a_{m}$, $a_e$ and $a_0$ are respectively the scale factors at the end of $m$-th phase transition, at the end of chain inflation and today. The scale factor ratio $a_m/a_e$ is equal to $\exp[-(N-m-1)\chi]$. If we also assume that the evolution from the end of chain inflation up to today has been adiabatic [\textit{i.e.} $S\propto a^3 g(T) T^3={\rm const}$], we have
\begin{equation}\label{aea0}
\frac{a_e}{a_0}\simeq 8\times 10^{-14}  {\left(\frac{100}{g_{\ast}}\right)}^{1/3} \left(\frac{1~{\rm GeV}}{T_e}\right).
\end{equation}
Thus
\begin{eqnarray}\label{fm2}
f_m= \frac{3\times 10^{-8}}{\chi}e^{-(N-m-1)\chi}{\left(\frac{g_{\ast}}{100}\right)}^{1/6}\left(\frac{T_e}{1~{\rm GeV}}\right),\\
\Omega h^2(f_m)=10^{-6}(1-e^{-\chi})^4 \chi^2
e^{-4(N-m-1)\chi}{\left(\frac{100}{g_{\ast}}\right)}^{1/3} .
\end{eqnarray}
The spectrum at other frequencies get redshifted such that the
profile given in eq.(\ref{GWprofile}) is preserved:
\begin{equation}\label{mth profile}
\Omega_m h^2(f)=\left\{
\begin{array}{ll}
\Omega h^2(f_m){\left(\frac{f}{f_{m}}\right)}^{2.8}\quad f\leq  f_{m} &\\
\Omega h^2(f_m){\left(\frac{f}{f_{m}}\right)}^{-1.8}\quad f\geq
f_{m}, &
\end{array}
\right.
\end{equation}
\begin{figure}[t]
\includegraphics[scale=0.35]{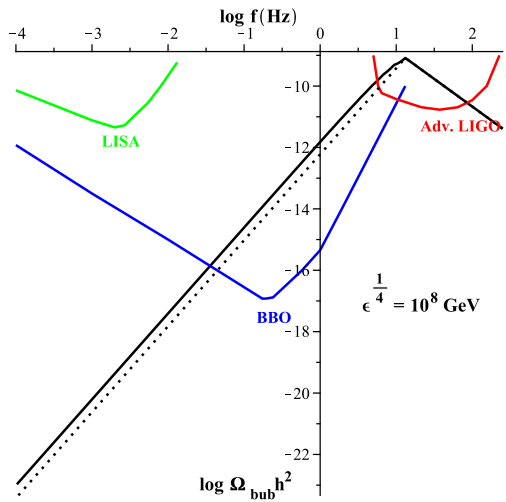}
 \caption{Solid line shows the gross features of the GW spectrum from bubble collisions in chain inflation for $\epsilon^{1/4}=10^8~{\rm GeV}$ and $\chi=\frac{1}{3}$, where $\epsilon$ is energy difference between vacua and $\chi$
 is number of $e$-folds per stage. For more detailed features near the peak,
 see Fig. 2. The dotted line
shows the GW spectrum generated from a single PT whose peak
frequency and  amplitude at the peak  coincide with those
from chain inflation; the small difference in the amplitudes at the low frequency tail is detectable.}\label{10to8}
\end{figure}

The total energy density in GW due to bubble collisions in chain inflation
is obtained by summing all the contributions from the multiple tunneling events,
\begin{equation}
\Omega_{\rm bub} h^2(f) = \sum_{m=1}^{N-1} \Omega_m h^2(f)
\end{equation}
and is plotted in Fig.(\ref{10to8}).  The frequency range plotted
is from $10^{-4}$ Hz  (roughly the minimum frequency attainable by
LISA and BBO) to $10^{2.5}$ Hz (roughly
the maximum sensitivity frequency of Advanced LIGO
\cite{advancedligo}). In the figure, we take $N=1000$ minima, $\epsilon^{1/4}=10^8$
GeV, and $\chi=\chi_c=1/3$ (the critical nucleation rate). For such parameters
the tip of the spectrum is observable in Advanced LIGO which is
sensitive within a frequency band of $1$-$1000$ Hz to stochastic
signals with $\Omega_{\rm GW}h^2\gtrsim 10^{-11}$. Part of the
low-frequency tail of the spectrum could be seen at BBO too, which
could probe GWs with much smaller amplitude, \textit{i.e.}
$\Omega_{\rm GW}h^2\gtrsim 10^{-17}$ for $f \sim (10^{-4}
- 10^1)$ Hz.
\begin{figure}[t]
\includegraphics[scale=0.35]{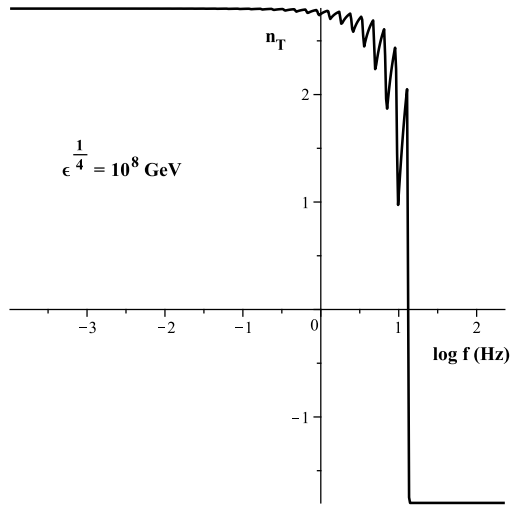}
 \caption{Tensor spectral index vs. frequency (blow-up of Fig. 1 near the peak frequency).
 For $f\ll f_{\rm peak}$ $n_T=-2.8$.  The wiggly structure in the spectrum
at $f\lesssim f_{\rm peak}$ is a unique signature for chain inflation.
For $f>f_{\rm peak}$, $n_T$ jumps to
$-1.8$. Alternatively, if we follow \cite{Huber:2007vva},  the tensor spectral index takes the value of $-1$ at the low-frequency tail.}\label{nT}
\end{figure}

One can see that the GW spectrum due to bubble collisions at frequencies smaller than the peak
frequency is blue. This is in contrast with slow-roll inflation
which can not produce a blue gravity spectrum on an extended range
of scales, unless one violates the null energy condition
$\dot{H}>0$. This is rooted in the sub-horizon mechanism of
generating the GWs from bubble-collision in chain inflation. As we
will see, chain inflation, like any other inflationary model, additionally
produces a red tensor spectrum at large scales through super-horizon
quantum fluctuations of the metric.

The most important contributions to the GW arise towards the very end of
inflation, during the last few PTs.
The peak frequency of the total spectrum is at
$f_{N-1}$ with amplitude $\Omega_{\rm bub}h^2 (f_{N-1})$.
The contribution from much earlier PTs --which have a
smaller peak frequency-- are (nearly)  invisible, because they are screened by the
low-frequency tail of the spectrum of subsequent PTs  due to
the redshift factor $e^{-4(N-m-1)\chi}$. That is why the number of
minima, $N$, does not change the final results as long as $N\chi\gg
1$.  However closer to the last bouts of chain inflation the effect
of redshift factor diminishes; thus the effect of PTs
that happened toward the end of chain inflation are in fact observable in the
spectrum as wiggles at $f\lesssim f_{\rm peak}$.

Focusing on the frequencies around the tip of the spectrum for chain inflation reveals a
richer structure: wiggles in the GW signal due to  different PT
giving contributions at different frequencies.  Since the resolution of Fig. 1 is inadequate to
show these wiggles,  Fig.  (\ref{nT}) illustrates
the tensor spectral index
$n_T\equiv d\ln \Omega_{\rm bub} h^2/2\pi d\ln f$ close to the peak
frequency $f_{N-1}$. This signature of wiggles near the peak from
chain inflation can be used to distinguish it from an arbitrary
single first order phase transition that happens to have the same
peak frequency and amplitude at the peak frequency.
Such wiggles can be a "smoking gun" for chain inflation.
\begin{figure}[t]
\includegraphics[scale=0.35]{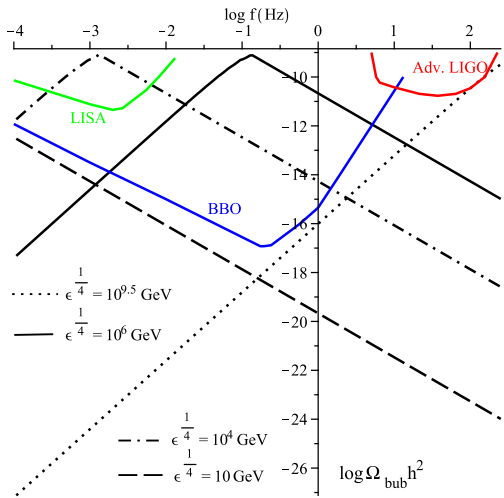}
 \caption{The spectrum of GWs produced from chain inflation for $\chi=1/3$ and for various values of $\epsilon$. For $10{\rm GeV}< \epsilon^{1/4}<10^{9.5} {\rm GeV}$ the signal is observable in some of the future GW probes.}\label{bubble-var-e}
\end{figure}
However, we caution that the details of the wiggles  have not yet been
computed precisely, e.g. we have not considered the effects of having the $m+1$th transition
start before the $m$th is complete.   Further work will be required to nail down the details.

There is a second way to differentiate multiple vs. single tunneling
events; i.e. to differentiate chain inflation from a single first
order PT with the same peak frequency and amplitude.
For both models to produce the same GW, the potential height $V_0$
of a single PT would have to be related to the energy difference
between vacua $\epsilon$ of chain inflation in the following way:
$V_0 \sim \epsilon / (1- {\rm exp}^{-\chi})^4$ (where $\chi$ is the
number of $e$-folds per PT in chain inflation).  The difference in
the overall amplitude of the spectrum can be seen (with poor
resolution) in Fig.1: matching both spectra at $f_{max}$, one finds
that the single PT has an overall amplitude that is {\it lower} than
that of chain inflationl by up to 3$\%$ at frequencies below the max
$f<f_{N-1}$.  The reason for this difference is the gradual decrease
of the tensor spectral index at frequencies close to but smaller
than $f_{N-1}$. This small difference in the amplitude at the low
frequency end can in principle be detected.

Three parameters in chain inflation determine the GW. The peak frequency in Eq (\ref{fm1}) depends on
$\chi$ (the number of $e$-folds per stage), $\epsilon$ (energy difference between any two vacua), and (very weakly) $N$ (the total number of minima); while the GW energy density in Eq (\ref{fm2}) depends on $\chi$ and (again very weakly) $N$.  Since the peak frequency and energy density scale exponentially with $-\chi N$ where
$\chi N >> 1$, the $N$-dependence can be ignored.  The largest GW amplitude is found
for the largest allowed value of $\chi$, i.e., $\chi = \chi_c =1/3$.  For the remainder of this paragraph
we investigate this case and vary $\epsilon$.  The GW amplitude remains the same
for all values of $\epsilon$, and the spectrum just shifts to a different frequency.
For $ 10^{7.5}~{\rm GeV}\lesssim \epsilon^{1/4} < 10^{9.5}~ {\rm
GeV}$, at least a part of the spectrum from
bubble collisions falls into the advanced
LIGO sensitivity region. For values of $10^{5.2}~{\rm GeV} \lesssim
\epsilon \lesssim 10^{7.5}~{\rm GeV}$, the peak frequency shifts to
the BBO region. Finally, for $10^{1.5}~{\rm GeV}
\lesssim\epsilon^{1/4} \lesssim 10^{5.2}~{\rm GeV} $, the peak
frequency shifts to LISA region. The gravitational wave signature of
chain inflation from bubble collision cannot be detected in these experiments  for
$\epsilon^{1/4}\lesssim 10~{\rm GeV}$ and $\chi = 1/3$. Alternatively, if we follow the results of \cite{Huber:2007vva}, scales as low as $\epsilon^{1/4}\sim {\rm few}~0.01 {\rm GeV}$ for $\chi = 1/3$
should be detectable. For $\chi=\frac{1}{3}$ and $\epsilon^{1/4}<{\rm few} ~{\rm eV}$, the amplitude of the spectrum from bubble collisions on large ($\sim$ horizon) scales becomes bigger than the observational bound from WMAP, $\Omega_{\rm GW}h^2<10^{-15}$. This puts a lower bound on the value of $\epsilon^{1/4}\gtrsim 10 ~{\rm eV}$, assuming that the nucleation rate takes the critical value, $\gamma=\frac{9}{4\pi}$.

\begin{figure}[t]
\includegraphics[scale=0.35]{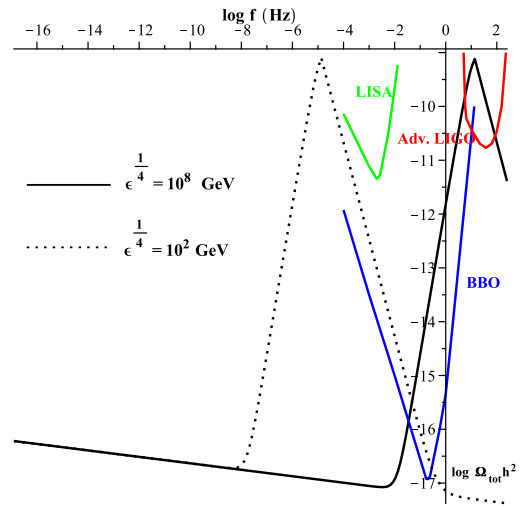}
 \caption{Total GW spectrum  produced during chain inflation from both bubble
 collisions and quantum fluctuations for $\chi = 1/3$, and for $\epsilon^{1/4}=100~{\rm GeV}$ and $\epsilon^{1/4}=10^8~{\rm GeV}$.}\label{total-spectra}
\end{figure}

One can investigate the dependence of the GW on the value of $\chi$.
We assumed so far that the nucleation rate is at the critical value so that
 $\chi=1/3$. For
smaller values of $\chi$, the peak frequency shifts to higher
values whereas the amplitude of gravitational waves decreases. Since
the peak frequency of the spectrum occurs at $f_{N-1}$, for a fixed
value of $\epsilon$, the peak frequency varies like
$\frac{1}{\chi(1-e^{-\chi})}$.  The dependence of the amplitude of the spectrum
on $\gamma$ is more involved, but as the amplitude at the peak
frequency is ${\rm few} \times \Omega_{\rm b}h^2(f_{N-1})$, one
would expect that the amplitude would depend on $\chi$ as $\chi^2
{(1-e^{-\chi})}^4$. One should note that for values of $\chi$ that
do not satisfy the condition $N\chi\gg 1$, the dependence of the
peak frequency and its amplitude on $\chi$ are respectively given as
$\frac{1-e^{-N\chi}}{\chi(1-e^{-\chi})}$ and  $\frac{\chi^2
{(1-e^{-\chi})}^4}{{(1-e^{-N\chi})}^4}$.

{\bf Spectrum of GWs from Quantum Fluctuations:} In addition to GW from bubble collisions, chain inflation, like
any other inflationary model, can produce a tensor spectrum from
quantum fluctuations of the space-time. The amplitude of such
quantum fluctuations is proportional to the overall amplitude of the
potential, $V$:
\begin{equation}\label{Qtensor}
P_T(k)=\frac{2}{3\pi^2}\frac{V}{M_{\rm P}^4}=\frac{8}{M_{\rm
P}^2}{\left(\frac{H}{2\pi}\right)}^2,
\end{equation}
where the expression should be calculated at the moment of
horizon-crossing. The relevant scale for comparison with CMB experiments  corresponds to $\sim N_e$
$e$-foldings before the end of inflation,
 where $N_e$ is the minimum number of e-foldings required to solve the horizon problem
 ($N_e \sim 60$ for GUT scale inflation); we take the corresponding potential height to
 be $V_\star$.
$\Omega_{\rm Q} h^2$ can be related to the
amplitude of primordial tensor perturbations through
\cite{Smith:2005mm}:
\begin{equation}\label{omega-r}
 \Omega_{\rm Q}h^2=A_{\rm GW} P_{T}=A_{\rm GW} r P_{S},
\end{equation}
where $A_{\rm GW}=2.74\times 10^{-6}
{\left(\frac{100}{g_{\ast}}\right)}^{1/3}$ and $P_S=2.45\times
10^{-9}$, at $k_{\star}=0.002 {\rm Mpc^{-1}}$ and WMAP+BAO+SN
constrain $r<0.22~(95\%)$C.L. \cite{Komatsu:2008hk}. In chain
inflation, the slope of the tensor power spectrum, $n_T$ relates to
$r$ through the relation
\begin{equation}\label{n_T-e}
 r=-\frac{8 n_T}{3}.
\end{equation}
which is different from the well-known consistency relation $r= - 8 n_T$ for
standard single-field slow-roll inflation (n.b., a different setup that violates the consistency relation is described
in \cite{Ashoorioon:2005ep}). The extra factor of $3$ in the denominator
of the right hand side (RHS) is due to the fact that in chain inflation, the scalar
spectrum is \cite{Chialva:2008zw}
\begin{equation}\label{Ps}
 P_S=\frac{H^2}{8\pi^2 M_{\rm P}^2 \frac{\varepsilon'}{3}}
\end{equation}
where $\varepsilon'$ is the equivalent of the slow-roll parameter as defined in
\cite{Chialva:2008zw}.
Thus for a given $r$, the tensor spectrum of chain inflation which
arises from quantum fluctuations is redder with respect to its
slow-roll counterpart. For example, for $r=0.01$, one finds $\Omega_{\rm
Q}h^2=6.58\times 10^{-17}$ at $k_{\star}=0.002 {\rm Mpc^{-1}}$
and $n_T=-3.75\times 10^{-3}$ for chain inflation. The total gravity
spectrum from chain inflation will be
\begin{equation}\label{total-spectrum}
 \Omega_{\rm tot}h^2=\Omega_{\rm bub} h^2+\Omega_{\rm Q}h^2.
\end{equation}
For values of $r$ detectable in the coming decades, the first term on the RHS dominates
on large scales while the second term dominates on small scales, as can be seen
in  fig.(\ref{total-spectra}).  The QUIET experiment \cite{Quiet} or the proposed CMBPOL \cite{inf-work-group}
have the capability of detecting tensor modes
down to at least $r = 0.01$ on Hubble scales,  $k=0.002~{\rm
Mpc^{-1}}$ ($f\sim 3.09\times 10^{-18}$~Hz).

The scale of the transition from dominance of quantum-induced GW to
bubble-induced GW depends on the values of $V_{\ast}$ (or,
equivalently, $r$), $\epsilon$ and $\chi$. For
$(r=0.01,\epsilon=10^8~{\rm GeV}, \,\,\, {\rm and} \,\,\,
\chi=\frac{1}{3})$, the transition occurs at $k_{t}\simeq 2.5\times
10^{12}~{\rm Mpc^{-1}}$  ($f\simeq10^{-2.4}$ Hz). Alternatively, for
$(r=0.01,\epsilon=10^2~{\rm GeV},$ and $\chi=\frac{1}{3})$, the
transition takes place at $k_t \simeq 4\times10^{6}~{\rm Mpc^{-1}}$;
see fig. (\ref{total-spectra}). For $\epsilon=10^2~{\rm GeV},$ and
$\chi=\frac{1}{3}$, quantum fluctuations will always dominate on
large scales as long as $r> 1.39\times 10^{-30}$ (\textit{i.e.}
$V_{\ast}> 8.73\times 10^{8} {\rm GeV}$) \footnote{In chain
inflation $r$ relates to the scale of inflation through $V_{\star}$
through $V_{\ast}^{1/4}=1.06\times 10^{16} {\rm GeV}
{\left(\frac{r}{0.03}\right)}^{1/4}$}. For a given $\chi$, the peak
frequency shifts to lower frequencies as $\epsilon$ decreases,
whereas the amplitude of the peak frequency remains almost constant.
For $\chi=\frac{1}{3}$ and $\epsilon^{1/4}<{\rm few} ~{\rm eV}$, the
amplitude of the spectrum from bubble collision becomes bigger than
the observational bound, $\Omega_{\rm GW}h^2<10^{-15}$. This puts a
lower bound on the value of $\epsilon^{1/4}\gtrsim 10 ~{\rm eV}$,
assuming that the nucleation rate takes the critical value,
$\gamma=\frac{9}{4\pi}$.

{\bf Conclusion:} In this letter, we calculated the profile of GWs
generated during chain inflation. Two mechanisms are responsible for
generating the signal; bubble collisions (which dominate on small
scales) and quantum fluctuations (which can dominate on large
scales). We found three observational signatures for chain
inflation. First, the signal could become detectable in upcoming
large frequency probes of gravity waves: Advanced LIGO, LISA, or
BBO. In fact, wiggles in the GW spectrum due to multiple phase
transitions may be a "smoking gun" at frequencies below the peak.
Second, the amplitude below the peak would be different from that of
a single tunneling model. Third, the consistency relation between
scalar and tensor modes from quantum fluctuations is different from
the standard one for slow-roll inflation and may be tested in CMB
experiments.


We also wish to comment on the connection with the landscape in
string theory. If inflation happened through a series of rapid
tunnelings in the string landscape, this work shows that future high
frequency probes of GW can shed light on the structure of the
landscape. Another possibility is that the many vacua in string
theory may be connected by a combination of  tunneling and rolling,
alternating with one another.  This case may generate the same GW
discussed in this paper due to bubble collisions, as long as several
of the last e-folds of inflation are due to tunneling. If there is
rolling near the beginning of inflation but tunneling near the end,
then the quantum fluctuations would produce the standard result
while the bubble collisions would be as described in this paper.  A
full analysis of the possibilities of alternating rolling and
tunneling warrants another paper. \\
\vspace{10mm}

\section*{Acknowledgments}
We are grateful to James Liu for useful discussions. We thank G.
Servant for sharing GW detector sensitivity curves.  A.A. is
partially supported by NSERC of Canada.  This work was supported in
part by the DOE under grant DOE-FG02-95ER40899 and by the Michigan
Center for Theoretical Physics.

\end{document}